\documentclass[fleqn,twoside]{article}
\usepackage{espcrc2}

\usepackage{graphicx,amssymb}
\usepackage[figuresright]{rotating}

\newcommand{\GeV}{~{\rm GeV}}

\newcommand{\AmS}{{\protect\the\textfont2
  A\kern-.1667em\lower.5ex\hbox{M}\kern-.125emS}}

\title{Direct extraction
of hadronic form factors from elastic-scattering data }

\author{E. Martynov\address[BITP]{Bogolyubov Institute for
        Theoretical  Physics,
        Metrologicheskaja 14b, UA-03143, Kiev, Ukraine}%
        \thanks{e-mail: martynov@bitp.kiev.ua},
        J.R. Cudell\address[ULG]{Institut de Physique, Universit\'{e} de Li\`{e}ge,
        4000 Li\`{e}ge, Belgium,}\thanks{e-mail: JR.Cudell@ulg.ac.be},
        A. Lengyel\address{Institute of Electron Physics, Universitetska 21, UA-88000
        Uzhgorod, Ukraine,} \thanks{e-mail: sasha@len.uzhgorod.ua}
        }

\begin{document}

\begin{abstract}
Non-forward elastic hadron-scattering data are collected and
analysed within the Regge approach. Through an analysis of the data
in small bins in $t$,
we have directly extracted the pomeron trajectory
and the hadronic form factors (or reggeon couplings). We found higher values
than usually used for the
intercept and for the slope of the soft pomeron
trajectory. The presence of zeros in $t$ for the effective
hadronic form factors is emphasised.

\vspace{1pc}
\end{abstract}

\maketitle

\section{Pomeron and reggeons at $t=0$}

A considerable effort \cite{COMPETE02} has recently been devoted to
the reproduction of soft data at $t=0$ through analytical fits based
on \( S \)-matrix theory. Three main forms for the pomeron, \( \ln
\frac{s}{s_{d}} \), \( \ln ^{2}\frac{s}{s_{t}}+C \), and simple
poles \( \left( \frac{s}{s_{1}}\right) ^{\alpha } \), work
reasonably well in the description of total cross sections at high
energy (\( \sqrt{s}>10 \)~GeV), however the simple-pole description
fails if the energy threshold is lowered to \( \sqrt{s}>5 \)~GeV, or
if the real part of the amplitude is included, whereas the
logarithmic forms achieve a good fit quality down to 5~GeV.

As shown in \cite{HARDPOM}, if a ``new'' singularity with higher
intercept $\alpha_{P}(0)\approx 1.45$ (unitarized at
$\sqrt{s}>100$~GeV) is added to the soft simple-pole-pomeron
contribution, all three forms describe the data almost equivalently.
On the other hand, the dispersion relations, with low-energy
corrections for the real part of amplitudes, lead also to improved and
comparable descriptions of the data at $t=0$ within these three
pomeron models \cite{IDR}. Thus,
one cannot conclude which kind of pomeron best agrees
with experiment on the sole basis of
the available data
on the total cross-sections, $\sigma_{t}(s)$ and $\rho(s)=\Re
eA(s,0)/\Im mA(s,0)$.

However, it is a very important task to cut down
the list of pomeron models as this is needed  to give unambiguous
predictions for high energies. We believe that, in order to make
progress in this direction, we have to extend the dataset,
considering also the data for non-forward elastic hadron scattering.

\section{Dataset for elastic scattering at $t\neq 0$}

For total cross sections and real parts of amplitudes at $t=0$, we
have now a standard and commonly accepted set of experimental
points \cite{PDG}. Unfortunately, such a standard dataset for
differential cross sections at $t\neq 0$ is not established yet.
However, the existing data are gathered in their original form, for
example in the Durham Database \cite{DDB}.

We have collected, checked, and uniquely formatted about 10000
experimental points on differential cross sections for $pp, \bar pp,
\pi^{\pm}p, K^{\pm}p$ elastic scattering in the energy region
$\sqrt{s}>4$~GeV and at all measured transferred momenta. The quality
of the data obtained in various experiments varies quite strongly.
For many sets, systematic errors are not given and for some others
it is not clear from the original papers whether the systematic
errors are included or not to the errors given in tables. Besides,
it is well known that the normalisations of different sets are not
in agreement, and are even sometimes in direct contradiction.

Taking into account these circumstances, we have excluded a few subsets (11)
out of a total of 200. These 11 subsets contradict even by eye the majority
of the other subsets. We do not give here the detailed description of
the data and individual references for the sets: this will be done in
our forthcoming full paper \cite{coming}.

\section{Analysis of the data}

A general goal of the analysis of the non-forward elastic-scattering
data is to compare various pomeron models on the basis of a standard
dataset, which should be as complete as possible. To simplify the
analysis at the first stage, we restrict ourselves to the
small-$|t|$ region, where the differential cross sections have a
smooth behaviour (dominated by single pomeron and reggeon exchange
terms). We analyse the data for elastic $pp, \bar pp, \pi^{\pm}p$
and $K^{\pm}p$ differential cross sections in the intervals
\begin{eqnarray}
\sqrt{s}&>&5\GeV \quad \mbox{for $pp$ and $\bar pp$},\nonumber\\
\sqrt{s}&>&6\GeV \quad \mbox{for $\pi p$ and $Kp$},
\end{eqnarray}
$$|t|_{min}=0.05\GeV^{2}<|t|<|t|_{max}=0.85\GeV^{2}$$
\noindent (There are some inconsistencies of the $\pi p$ and $Kp$ data at $5\GeV
<\sqrt{s}< 6\GeV$, hence the exclusion of this region).
To analyse the data, we considered
a simplified parametrisation for the amplitudes, and then used it
in the so-called method of overlapping bins.

\noindent{\bf Amplitudes:} they were written in the form
\begin{equation}
A_{ab}^{\bar ab}(s,t)=P_{ab}(s,t)+R_{ab}^{+}(s,t)\pm
R_{ab}^{-}(s,t),
\end{equation}
where the pomeron term $P_{ab}$, the crossing-even term $R_{ab}^{+}$
and the crossing-odd term $R_{ab}^{-}$ are
\begin{equation}
P_{ab}(s,t)=
\frac{g_{ab}(-is_{ab}/s_{0})^{\alpha_{P}}}{-\sin(\pi\alpha_{P}/2)},
\end{equation}
\begin{equation}
R_{ab}^{+}(s,t)=
\frac{g_{ab}^{+}(-is_{ab}/s_{0})^{\alpha_{+}}}{-\sin(\pi\alpha_{+}/2)},
\end{equation}
\begin{equation}
R_{ab}^{-}(s,t)=
\frac{ig_{ab}^{-}(-is_{ab}/s_{0})^{\alpha_{-}}}{\cos(\pi\alpha_{-}/2)},
\end{equation}
with $s_{0}=1$~GeV$^{2}$ and $s_{ab}=s-m_{a}^{2}-m_{b}^{2}+t/2
\propto f(t)\cos\theta_{t}$,
$\cos\theta_{t}$ being the scattering angle in the
$t$-channel.
We consider the crossing-even and crossing-odd reggeons as
effectively describing exchange-degenerate $f-a_{2}$ and $\omega-\rho$
contributions. The couplings $g$ and trajectories $\alpha$ are the
constants determined from the fit in each $t$-bin.

\noindent{\bf Overlapping bins:} Let us define the elementary bin:
$\tau<-t<\tau+\Delta t$. If the $s$-dependence of the above
amplitudes is roughly correct, we can determine the values of the
trajectories and couplings in a given bin, provided $\Delta t$ is
small enough. Performing such fit for $\tau=|t|_{min}+k\cdot\delta
t, k=1,2,...,\, k_{max}$, we scan all data from $|t|_{min}$ up to
$|t|_{max}$ and extract an effective $t$-dependence of trajectories
and form factors or couplings. A few remarks regarding the method
are to be noted:
\begin{enumerate}
\item
\noindent The length of the bin should be not too big because
of the simplified parametrisation and not too small in order to contain
a reasonable number of points for each process. We take \break$\Delta
t=0.025 \GeV^{2}$.\vglue 6pt
\item
\noindent We have chosen $N_{min}=4$ as the minimal number of points for
each process, thus taking into account only
bins with 24 points or more.\vglue 6pt
\item
\noindent We take for the shift from bin to bin \break$\delta t=0.01\GeV^{2}$.
Wide variations of $\Delta t$ and $\delta t$ do not change
the conclusions.\vglue 6pt
\item
\noindent The fact that $d\sigma/dt$ decreases almost exponentially
with $|t|$ is important even for small $\Delta t$.
In each bin, we have written all couplings in the form
$g\to g(t)=g\exp(b(t-\tau))$. The constants $g$ (effective couplings
at the first point ($t=\tau$) of the corresponding bin) are
discussed below.\vglue 6pt
\end{enumerate}
\noindent {\bf Reggeon (``$f$'' and ``$\omega$'') trajectories:}
we are faced with the problem of the crossing-even and crossing-odd
reggeon terms. Let us to call them as ``$f$'' and ``$\omega$''
contributions. They are badly determined from the fit. The obtained
values of trajectories, $\alpha_{+}$ and $\alpha_{-}$, as well as
$g^{\pm}$, if left free, have a big dispersion from bin to bin. This
is caused mainly by the inconsistencies in normalisation of the many
subsets. To restrict this randomness, we used the intercepts and
slopes of $f$ and $\omega$ trajectories from the fits at $t=0$ and
from the spectroscopic data \cite{COMPETE02,IDR,PDG,DEGENER}.
Namely, we fixed the $\omega$ trajectory using
$\alpha_{-}=0.445+0.908\ t$. For the $f$ trajectory we considered
two variants, ``maximal'' $\alpha_{+}=0.697+0.801\ t$ and
``minimal'' $\alpha_{+}=0.615+0.820\ t$.

\section{The results}
\begin{figure}[h]
\begin{center}
\includegraphics*[scale=0.5]{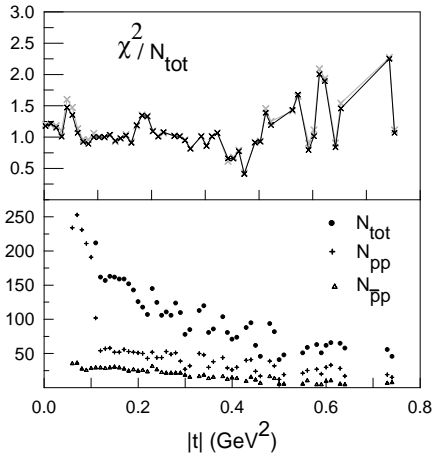}\vglue -1cm
\caption{$\chi^{2}$ per point
and number of points (for $pp$ and $\bar
pp$) for each bin.} \label{fig:chi}
\end{center}
~\vglue -1cm
\end{figure}

\begin{figure}[h]
\begin{center}
\includegraphics*[width=6.6cm, height=4cm]{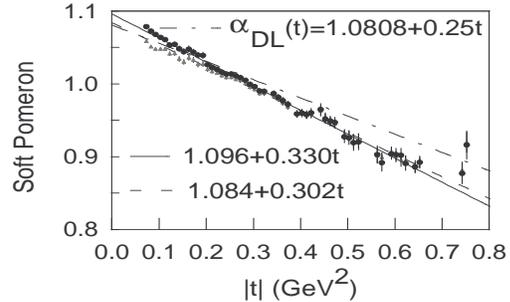}\vglue -1cm
\caption{Fitted pomeron trajectory. The dashed-dotted line
corresponds to the standard pomeron trajectory \cite{DDLN}, and the
two other lines to our fit to the extracted trajectory for minimal
and maximal $f$ trajectories.} \label{fig:traj}
\end{center}
~\vglue -1cm
\end{figure}

\begin{figure}[h]
\begin{center}
\includegraphics*[width=7.5cm, height=3.2cm]{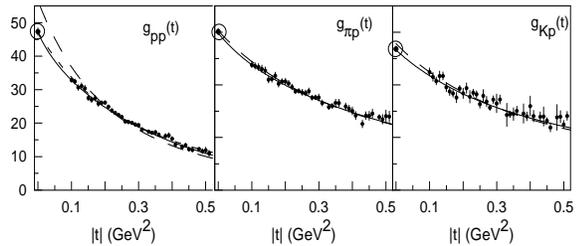}\vglue -0.5cm
\caption{Pomeron couplings, points inside circles are added
from the fit at $t=0$. The long-dashed line corresponds to the
parametrisation (\ref{DDLN}), the short-dashed line to a dipole form
(\ref{dipol}), and the solid line to the
sum of two exponentials (\ref{expo}).} \label{fig:pcoupl}
\end{center}
~\vglue -1cm
\end{figure}

\begin{figure}[h]
\begin{center}
~\vglue -1cm\includegraphics[width=7.5cm, height=3.cm]{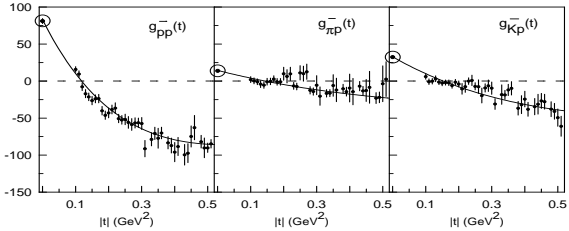}\vglue -1.0cm
\caption{$\omega$-reggeon couplings, points inside circles
are added from the fit at $t=0$.} \label{fig:ocoupl}
\end{center}
~\vglue -1cm
\end{figure}

\begin{figure}[h]
\begin{center}
\includegraphics*[width=7.5cm, height=3.cm]{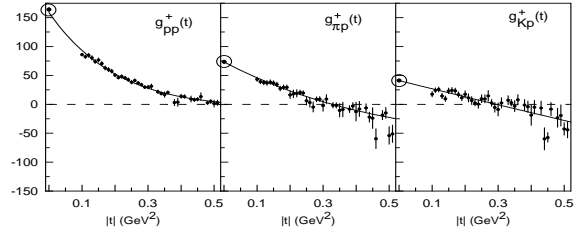}\vglue -0.5cm
\caption{$f$-reggeon couplings, points inside circles are
added from the fit at $t=0$.} \label{fig:fcoupl}
\end{center}
~\vglue -1cm
\end{figure}

The results obtained in the bin-analysis of the differential cross
sections are shown in Figs \ref{fig:chi}-\ref{fig:fcoupl}. The black
(resp. grey) symbols in Figs. \ref{fig:chi}, \ref{fig:traj}
correspond to the maximal (resp. minimal) $f$ trajectory.
\vspace{-0.5cm}

A good description of the data ($\chi^{2}/N_{tot}\approx 1$, see
Fig.~\ref{fig:traj}) is achieved in the interval $0.1
\GeV^{2}\leq |t|\leq 0.5\GeV^{2}$. We can thus treat this
interval as the first diffraction cone.

{\noindent\bf Pomeron trajectory:} one can see from Fig.~\ref{fig:traj}
that data do not confirm the
``standard'' values of the soft pomeron trajectory \cite{DDLN}. We indeed obtain
\begin{eqnarray}
\alpha_{P}(0)&=&1.084-1.096 \\
\alpha_{P}'&=&0.3 - 0.33 \GeV^{-2}
\end{eqnarray}
instead of $\alpha_{P}(0)=1.0808$ and $\alpha_{P}'=0.25\GeV^{-2}$.

\noindent{\bf Pomeron couplings} Fig.~\ref{fig:pcoupl} shows that,
in the $pp$ and
$\bar p$ case, the dependence of $g_{ab}$ on $t$ is not described perfectly
($\chi^{2}/N_{dof}=2.92$) by the expression
\begin{equation} g_{ab}(t)=g_{ab}\left
[\frac{1-t/1.269}{(1-t/3.519)(1-t/0.71)^{2}}\right ]^{2}
\label{DDLN}
\end{equation}
given in \cite{DDLN}. They are much better
described ($\chi^{2}/N_{dof}=0.95$) by a simple dipole form
\begin{equation}
g_{ab}\left [ \frac{1}{(1-t/3.519)(1-t/0.661)}\right ]^{2}
\label{dipol}
\end{equation}
or by the sum of two exponentials ($\chi^{2}/N_{dof}=0.92$)
\begin{equation}
g_{ab}\left [ 0.08\exp(12.69~t)+0.92\exp(1.27~t)\right ]^{2}.
\label{expo}
\end{equation}

\noindent {\bf Reggeon couplings:} the effective couplings of $f$
and $\omega$ reggeons both have at least one zero. (Figs.
\ref{fig:ocoupl} - \ref{fig:fcoupl}). For the $\omega$ contribution,
this is a well-known feature, related to the crossover effect
(differential cross sections of $ab$ and $\bar ab$ cross at small
$|t|$) discovered a long time ago in several experiments. However
the zeros in the $f$-reggeon couplings are highly unexpected.

The detailed analysis of the results for the $t$-dependence of the
effective couplings will be performed in a forthcoming
full paper \cite{coming}.
Here we
give only some properties of the derived couplings
in the case of the maximal $f$ trajectory.

\noindent{\bf Crossing-odd couplings:} Fig.~\ref{fig:ocoupl} shows
their extracted values and a fit to a single exponential multiplied by
the factor
$1+t/t_{\omega}$. This leads to the following values for the zeros (in
GeV$^{2}$):
$$
t_{\omega}=-0.12 \, (pp) \quad =-0.14 \, (\pi p) \quad =-0.15 \,
(Kp)
$$

\noindent {\bf Crossing-even reggeon couplings:} Fig.~\ref{fig:fcoupl} shows
that the
same procedure as for the $\omega$ couplings gives zeros of the
$f$ couplings at
$$
t_{f}=-0.6 \, (pp) \quad =-0.33 \, (\pi p) \quad =-0.29 \, (Kp)
$$
again in GeV$^2$.

As a conclusion to this short note, we would like to point out that the
interpretation of the zeros in the form factors is not yet clear. Our
attempts to perform a global fit of the various analytical models to
the data failed when double re-scatterings (even with
arbitrary strength) were included. Only the explicit account
of zeros allows to describe well the full set of data in the first
cone region, however it leads to low slopes for the $f$ and $\omega$
trajectories. A more thorough investigation of the problem is in progress.

\end{document}